\documentclass[twoside]{article}
\usepackage{saint-malo2011}
\usepackage{url}

\toappear{1} 

\begin{document}

\parindent=0pt

\bibliographystyle{plain}

\title{\Large\bf A Modeling Approach based on UML/MARTE for GPU Architecture}
\shorttitle{A Modeling Approach based on UML/MARTE for GPU Architecture}

\author{Antonio Wendell de O. Rodrigues, Frédéric Guyomarc'h and Jean-Luc Dekeyser}

\address{LIFL - USTL\\\small{INRIA Lille Nord Europe} - 59650\\Villeneuve d'Ascq - France\\wendell.rodrigues@inria.fr}

\date{February, 8 2011} 

\maketitle

\Resume{Nowadays, the High Performance Computing is part of the context of embedded systems. Graphics Processing Units (GPUs) are more and more used in acceleration of the most part of algorithms and applications. Over the past years, not many efforts have been done to describe abstractions of applications in relation to their target architectures. Thus, when developers need to associate applications and GPUs, for example, they find difficulty and prefer using API for these architectures. This paper presents a metamodel extension for MARTE profile and a model for GPU architectures. The main goal is to specify the task and data allocation in the memory hierarchy of these architectures. The results show that this approach will help to generate code for GPUs based on model transformations using Model Driven Engineering (MDE).}
\MotsCles{GPU, Embedded Systems, MDE, Code Generation}

\section{Introduction}
Over the past years, software researchers and developers have been creating abstractions
that help them program in terms of their design
intent rather than the underlying architectures, e.g., CPU, memory,
and network devices. Moreover, they shield themselves from the complexities
of these architectures.

Today, developers need to specify their software based on used platform architectures. Model Driven Architecture (MDA)\cite{Bezivin} is a framework proposed by Object Management Group (OMG) for software development. This framework is driven by models in different abstraction levels. Approaches based on MDA have been used as a solution to accelerate the embedded system design. In this solution, initially a system is modeled using a Platform-Independent Model (PIM). Then, using transformation languages, this model is transformed in a Platform-Specific Model (PSM).

In order to define these models, the OMG proposes Meta Object Facility (MOF)\cite{MOF} and Unified Modeling Language (UML) as standard for modeling and meta-modeling. Furthermore, the OMG adds the profile concept which provides a generic extension mechanism for customizing UML models for particular domains and platforms. Modeling and Analysis of Real-time and Embedded systems (MARTE)\cite{marte10} is a profile that adds capabilities to UML for model-driven development of Real Time and Embedded Systems (RTES).

Lately, a number of researches in High Performance Computing (HPC) have led to use of Graphics Processing Unit (GPU). GPU is a manycore processor attached to a graphics card dedicated to calculating floating point operations. Initially, the GPU was dedicated to graphics operations in personal and desktop computers. But, its processing power allowed executing other algorithms. This was called General Purpose GPU (GPGPU). These processors have been progressively part of embedded systems \cite{ES}. Thus, in order to optimize their application performance, embedded systems can use the processing power of GPUs. In this paper, we present a metamodel for designing of GPU characteristics, and a model for a specific GPU architecture. Both metamodel and model are part of \textit{Gaspard2}\cite{gaspard2}, which is an environment for development of RTES and Intensive Signal Processing (ISP) applications. \textit{Gaspard2} is the flagship project of DaRT team. 

We divide this paper in eight sections. The section 2 shows the main aspects of model driven engineering. The next one presents the purpose of MARTE profile. The GPU architecture is described in the section 4. Then, in the next section we show our metamodel approach. The section 6 depicts an example model for devices based on GPU. Finally, in the section 7 and 8, we presents the environment of development with a case study and conclusions respectively.
in
\section{Model Driven Engineering}
Model Driven Engineering (MDE) is a software development methodology which focuses on creating models, or abstractions, closer to some particular domain concepts rather than computing (or algorithmic) concepts. It is meant to increase productivity by maximizing compatibility between systems, simplifying the process of design, and promoting communication between individuals and teams working on the system. The best known MDE initiative is the MDA, which is a registered trademark of OMG.

\begin{figure}[ht]
\centering
\includegraphics[width=.7\textwidth]{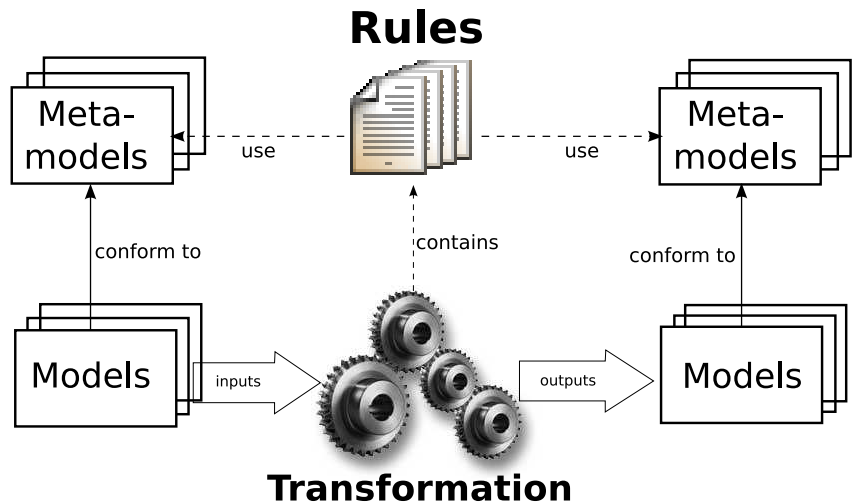}
\caption{Model Transformation Illustration}
\label{modeltransf}
\end{figure}

In MDE, complex systems can be easily understood thanks to abstract and simplified
representations: models. Graphical representations of models considerably
facilitate the comprehension of a given model. The UML language has been often used
for such graphical representations since its normalization in 1997. A model
highlights the intention of a system without describing the implementation
details. MDE is definitely oriented towards the modeling of software engineering systems. The
resulting models must be comprehensive and interpretable by computer.  In addition, in MDE, we can use Domain-Specific Modeling Language (DSL) to model systems with peculiar characteristics.
MDE also covers the code generation, which puts a model in concrete form. In this
way, MDE stands apart from the others methodologies based on models. The next subsections detail the major aspects of MDE that are model, metamodel and model
transformations. 

\subsection{Model}
A model is an abstraction of the reality. Models are composed of concepts and
relations. The concepts represent an abstraction of objects and
relations represent the links between the objects. In addition, models can be graphically
observed from different points of view (views in MDE), which highlight specific
aspects of the reality.

Applications and architectures of embedded systems have clearly identified elements (objects) such as
data parallel tasks, data dependencies, multidimensional data arrays, and architecture parts. The
abstraction of each element corresponds to a concept in a model, the dependencies
between these elements are represented by relations. Models can represent
abstract descriptions of these applications and thus, it helps to specify and modify them since each concept and relation are clearly identified. Moreover,
views can help to represent and document models by highlighting the relevant concepts and relations according to a particular
purpose.

\subsection{Metamodel}
A metamodel gathers the set of concepts and relations between the concepts
used to describe a model, i.e., the reality according to a particular purpose
(a given abstraction level for instance). Then a model conforms to a
metamodel which specifies a modeling structure. 
In the other words, a metamodel defines the syntax of its models, like a grammar defines its language.
Consequently, a metamodel can state the set of necessary concepts and relations to
represent the applications and architectures of embedded systems at a given abstraction level.
A model always conforms to a metamodel. This relation is called conformance. The conformance relation has a different nature than the representation relation between a model and its system. A metamodel does not represent a model (that could be considered a system), but only the concepts and relationships that may be created. Additionally, a metametamodel defines the syntax of its metamodel, then a metamodel conforms to a metametamodel.

\subsection{Model Transformations}
In MDE, a model transformation is a compilation process which transforms
a source model into a target model. The source and
the target models are respectively conformed to the source and the target
metamodels (fig. \ref{modeltransf}). A model transformation relies on a set of rules. Each rule clearly
identifies concepts in the source and the target metamodels. Such decomposition
facilitates the extension and the maintainability of a compilation process: new
rules extend the compilation process and each rule can be modified independently
from the others.
The rules are specified with languages. The language may be imperative:
it describes how a rule is executed; it can be declarative, it describes what is
created by the rules. Declarative languages are often used in MDE because the
rules objectives can be specified independently from the execution. A graphical
representation is a good approach for representing the rules expressed in a
declarative language.

\section{MARTE Profile}
\begin{figure}[ht]
\centering
\includegraphics[width=.7\textwidth]{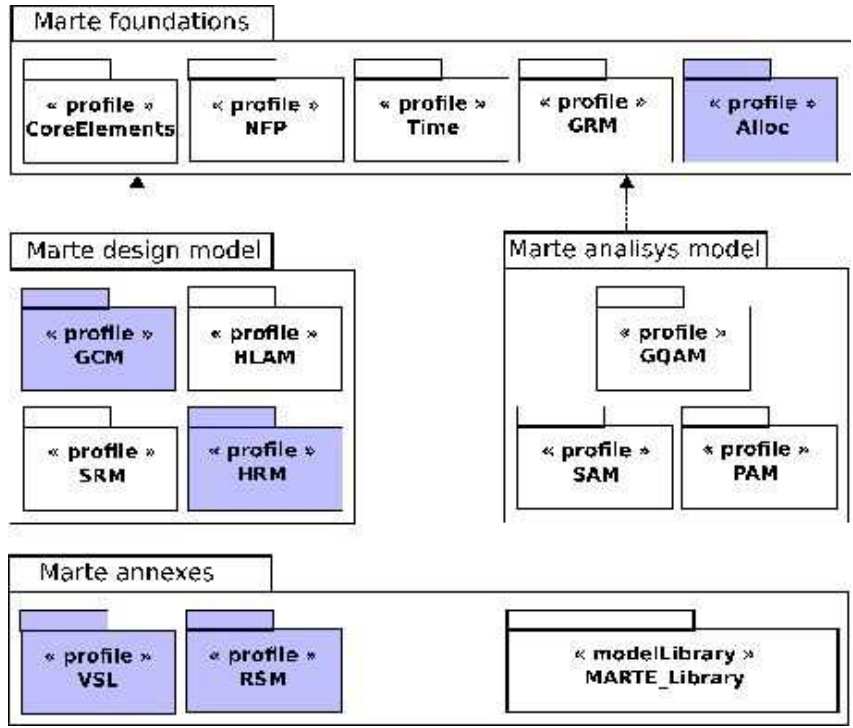}
\caption{Architecture of the MARTE profile}
\label{martearchi}
\end{figure}
The UML profile for MARTE (or MARTE profile) extends the possibilities for modeling of application and architecture and their relations. In addition, MARTE allows extending the performance analysis and task scheduling based on target platform architecture.

MARTE consists in defining foundations for model-based description of real time and embedded systems. These core
concepts are then refined for both modeling and analyzing concerns. Modeling parts provide support required from
specification to detailed design of real-time and embedded characteristics of systems. MARTE concerns also model-based
analysis. In this sense, the intent is not to define new techniques for analyzing real-time and embedded systems, but to
support them. Hence, it provides facilities to annotate models with information required to perform specific analysis.
Especially, MARTE focuses on performance and schedulability analysis. But, it defines also a general analysis framework
which intends to refine/specialize any other kind of analysis. Among others, the benefits of using this profile are thus:
\begin{itemize}
    \item providing a common way of modeling both hardware and software aspects of a RTES in order to improve
       communication between developers;
    \item enabling interoperability between development tools used for specification, design, verification, code generation, etc.;
    \item fostering the construction of models that may be used to make quantitative predictions regarding real-time and
       embedded features of systems taking into account both hardware and software characteristics.
\end{itemize} 
The fig. \ref{martearchi} shows the MARTE profile architecture. Moreover, we highlight some of our contributions within DaRT team. Over the last years, the DaRT team has contributed to specify the \textit{Allocation Modeling} (Alloc), \textit{Generic Component Modeling} (GCM), \textit{Hardware Resource Modeling} (HRM), \textit{Value Specification Language} (VSL), and \textit{Repetitive Structure Modeling} (RSM). In this paper, we focus on the Alloc, HRM and RSM components. In addition, we purpose a \textit{memory mapping} metamodel in order to describe the data allocations in the memory.

\begin{figure}[ht]
\centering
\includegraphics[width=.7\textwidth]{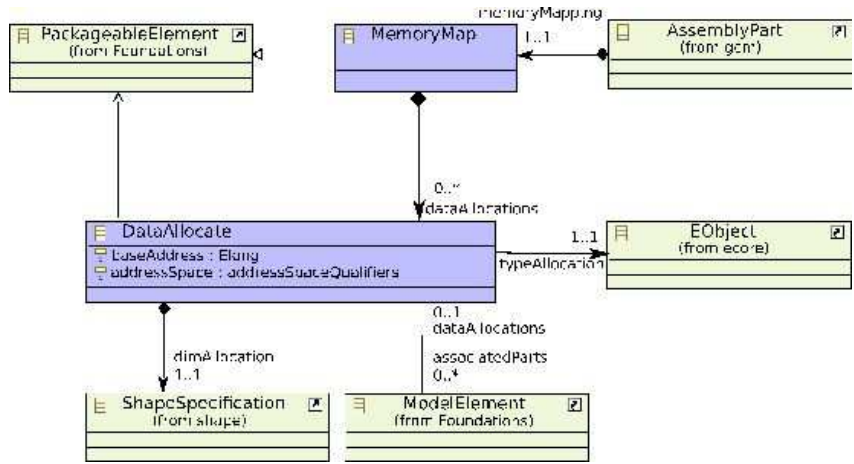}
\caption{Memory Mapping Metamodel Extension for MARTE}
\label{mmmetamodel}
\end{figure}

\section{Graphics Processing Unit}
The GPU devotes more transistors to data processing rather than data caching and flow control. This is the reason why the GPU is specialized for compute intensive. NVIDIA GPUs, more precisely, are composed of array of Streaming Multiprocessors (SM or Compute Units), which is equipped with 64 scalar cores (the SP, Streaming Processors or Processing Elements), 16834 32-bit registers, and 48KB of high-bandwidth low-latency memory shared for up to 1024 co-resident threads (or work-items). GPUs such as the NVIDIA GeForce GTX 480 contain 15 Streaming Multiprocessors, each of which supports up to 1024 co-resident threads, so 30K threads can be created for a certain task. In addition, each multiprocessor executes groups, called warps, of 32 threads simultaneously.
NVIDIA's actual CUDA architecture, code-named Fermi, has features for general-purpose computing. Fundamentally, Fermi processors are still graphics processors, not general-purpose processors. The system still needs a host CPU to run the operating system, supervise the GPU, provide access to main memory, present a user interface, and perform everyday tasks that have little or no data-level parallelism.

\subsection{Memory Architecture}
In the NVIDIA GPU memory hierarchy, there are per-thread local, per-block shared, and device memory which comprehend global, constant, and texture memories. Shared memory can be only accessed by threads in the same block. The shared memory space is much faster than the local and global memory spaces due to placement on chip. But, except for Fermi, only 16KB of shared memory are available on each SM.

\subsection{Programming Model}
Compute Unified Device Architecture (CUDA) is a C language extension developed by NVIDIA to facilitate writing programs on GPUs. This extension allows the programmer to define C functions, called kernels, that, when called, are executed N times in parallel by N different CUDA threads, as opposed to only once like regular C functions. One of the main features of CUDA is the provision of a Linear Algebra library (CuBLAS) and a Fast Fourier Transform library(CuFFT) \cite{cuda}.
Khronos Group has released the specification to OpenCL (currently 1.1). OpenCL (Open Computing Language) is the an open, royalty-free standard for general-purpose parallel programming of heterogeneous systems. It provides an uniform programming environment for software developers to write efficient, portable code for high-performance computing servers, desktop computer systems and handheld devices using a diverse mix of multi-core CPUs, GPUs, Cell-type architectures and other parallel processors such as DSPs. The OpenCL has some similarities with CUDA programming model \cite{opencl10}.

\section{MARTE Metamodel Extensions}
\begin{figure}[ht]
\centering
\includegraphics[width=.7\textwidth]{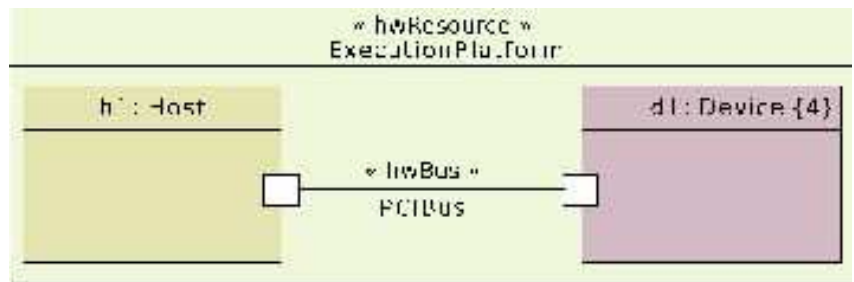}
\caption{Host and Device Architecture}
\label{h1gpu}
\end{figure}

The UML Tools provide a standard representation of UML diagrams and concepts. UML also allows to extend these concepts by the use of profiles and stereotypes. Actually, representation of these stereotypes can be customized in a limited way. MARTE profile provides us stereotypes to describe new elements in an UML model. Stereotypes are one of three types of extensibility mechanisms in UML. They allow designers to extend the vocabulary of UML in order to create new model elements, derived from existing ones, but that have specific properties that are suitable for a particular problem domain or otherwise specialized usage. There are two ways for defining UML-based Domain-Specific Modeling Languages (DSL). In the first one, DSL can be defined as a UML profile, such as MARTE profile, in a lighweight way, using stereotypes and tagged values. In the second way, the Meta Object Facility (MOF) can be used to either extend the UML metamodel, or to direcly define the metamodel without dependency on UML. We chose the second one for extending MARTE. The purposed metamodel defines an abstract syntax for our DSL.

\begin{figure}[ht]
\centering
\includegraphics[width=.7\textwidth]{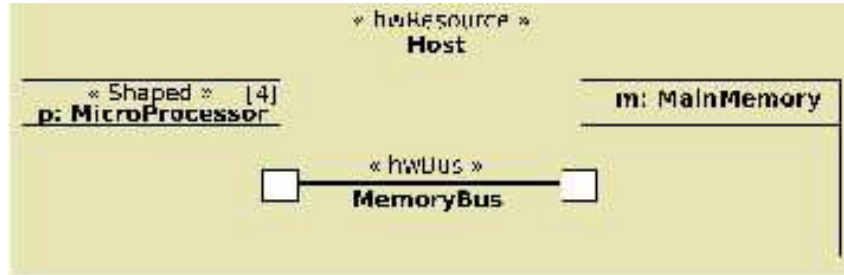}
\caption{Host Components}
\label{h2gpu}
\end{figure}

The DaRT team has worked on metamodel extensions for MARTE. Among these extensions, we specialized an extension for memory concepts in modeling GPU architecture. The fig. \ref{mmmetamodel} shows the abstract syntax that defines memory mapping concepts. Two new classes were created to add concepts for data allocation over a memory map: \textit{MemoryMap} and \textit{DataAllocate}. An \textit{AssemblyPart}, generally an instance of memory component in the defined model, should have one \textit{memoryMapping} of \textit{MemoryMap} type which is composed or not by \textit{dataAllocations} of the \textit{DataAllocate} type. Each \textit{dataAllocation} has its data scopes. This scope (\textbf{\textit{spaceAddress}}) is defined by \textbf{\textit{addressSpaceQualifiers = \{global,constant,local,private\}}} such as speficied in GPU memory hierarchy. Additionally, the \textit{dataAllocation} has:
\begin{enumerate}
\item \textbf{\textit{baseAddress}}: of the long integer data type and it specifies the base reference value or the base pointer of the variable;
\item \textbf{\textit{dimAllocation}}: this information comes from \textit{flowPorts shape} in the model and it defines the allocation size;
\item \textbf{\textit{associatedParts}}: it lists all the elements in the model that use the same allocation;
\item \textbf{\textit{typeAllocation}}: this information comes from \textit{flowPorts type} in the model and it defines the variable datatype.
\end{enumerate}

\begin{figure}[ht]
\centering
\includegraphics[width=.7\textwidth]{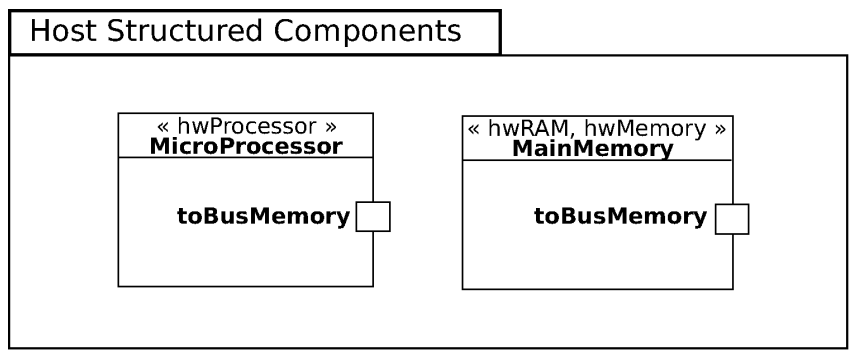}
\caption{Host Structured Components Package}
\label{hostcomp}
\end{figure}

This metamodel, also known as \textit{MemoryMapping} metamodel extension for MARTE, allows us to specify variables and their attributes. We can obtain the information from \textit{flowPort} elements which are specified in the application model.

\section{GPU Architecture Models}

\subsection{Platform Model}
\begin{figure}
\centering
\includegraphics[width=.45\textwidth]{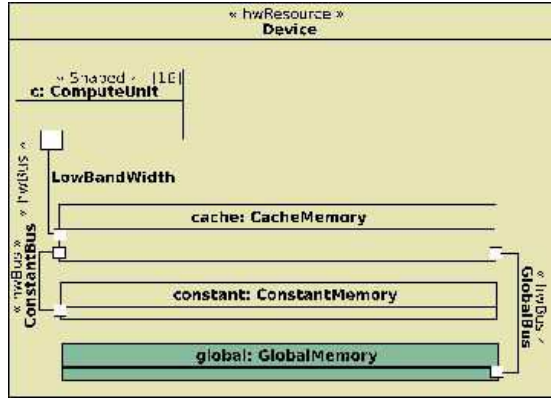}
\caption{Device Internal Model}
\label{h3gpu}
\end{figure}

In order to run applications on GPUs, developers should take into account that GPUs do not work alone. GPUs are coprocessors that need a host. Nowadays, CPUs with applications based on C language are the most suitable host for GPUs. The fig. \ref{h1gpu} depicts the host and device modeling. A databus, generally a \textit{pcibus}, links host and device. The host uses this bus to communicate with the device for threads(work-items) launches and data copies. In our model, the host is divided in two component instances of \textit{MicroProcessor} and \textit{MainMemory}. The first one is a component that represents the processor and it has a \textit{hwProcessor} stereotype (fig. \ref{hostcomp}) from MARTE profile. This stereotype allows specifying processor details, such as \texttt{speedFactor}, \texttt{frequency} or \texttt{isActivate} status. In addition, the \textit{MicroProcessor} instance can be stereotyped with \textit{Shaped}. With this stereotype, the instance can provide multiplicity informations. The fig. \ref{h2gpu} shows a shaped processor with a single dimension multiplicity which is equal to 4. The \textit{Shaped} stereotype has an important role in application task distribution over processors. The \textit{hwProcessor} and \textit{Shaped} stereotypes are described in HRM and RSM packages of MARTE.

\begin{figure}
\centering
\includegraphics[width=.45\textwidth]{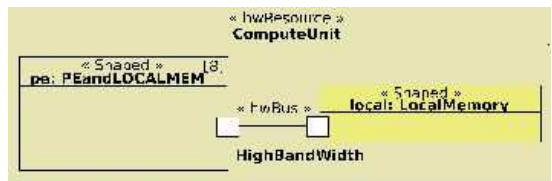}
\caption{Internal Structure of Compute Units}
\label{h4gpu}
\end{figure}

The device in the architecture model is the GPU. As our intent is modeling GPU architecture, we add more information about the interation between components of its structure. Since CUDA is proprietary, we opted for OpenCL standard \cite{opencl10}. In OpenCL, a device is divided into one or more compute units (CUs) which are further divided into one or more processing elements (PEs). Computations on a device occur within the processing elements. In the presented model in fig. \ref{h3gpu}, we have an instance \textit{c} of the \textit{ComputeUnit} type and \textit{Shaped} value is equal to 16. Each \textit{ComputeUnit } is composed of 8 \textit{ProcessingElement} as described in fig. \ref{h4gpu} and fig. \ref{h5gpu}.

\subsection{Memory Model}
\begin{figure}
\centering
\includegraphics[width=.45\textwidth]{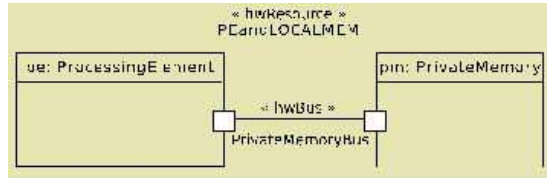}
\caption{Internal Structure of Processing Elements}
\label{h5gpu}
\end{figure}

In order to obtain a clear and useful model, we specified a simple memory structure in the host side. The host has one single shared memory component. This component is \textit{hwMemory} or \textit{hwRAM} stereotyped, thus we can map \textit{flowPorts} over this memory instance and obtain allocations for application variables.

However, the device has more complex memory structure. There is a hierarchical memory composition that allows different visibility and bandwidth.

\begin{figure}
\centering
\includegraphics[width=.45\textwidth]{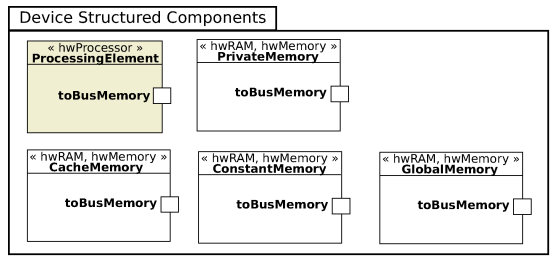}
\caption{Device Components}
\label{devicecomp}
\end{figure}

\begin{enumerate}
\item \textbf{\textit{Global Memory}}: work-items(threads on NVIDIA definition) can read from or write to any element of a memory object.
Reads and writes to global memory may be cached depending on the capabilities of the
device;
\item \textbf{\textit{Constant Memory}}: a region that remains constant during the
execution of a kernel. The host allocates and initializes memory objects placed into
constant memory;
\item \textbf{\textit{Local Memory}}: a memory region local to a compute unit. This memory region can be
used to allocate variables that are shared by all work-items placed on the same compute unit. It may be
implemented as dedicated regions of memory on the device;
\item \textbf{\textit{Private Memory}}: a region of memory private to a work-item. Variables defined in one
work-item's private memory are not visible to another work-item. Each processing element has its private memory as illustrated in the fig. \ref{h5gpu}
\end{enumerate}

All base components of device (including processor and different memory blocks) are presented in fig. \ref{devicecomp}.

Since our model has all memory hierarchy levels of the device, developers can identify which elements from application model can be placed in one of these levels. The next subsection depicts the variable allocations which are the aim of this placement. The aim is to construct an structured model that allows allocating data in the memory according to the application needs.

\subsection{Variable Allocations}
A critical problem in application modeling based on MDE is to manage the memory allocation in the target platform. MARTE profile adds the \textit{flowPort} stereotype to UML \textit{port} element. The main attribute that is aggregated on port element is the \textit{direction}, which allows to define if the port is input, output, or bidirectional. This information contributes for deciding which elements are read-only. The other information to define a variable are provided from elements described in \textit{MemoryMapping} metamodel (section V).

\begin{figure}
\centering
\includegraphics[width=.7\textwidth]{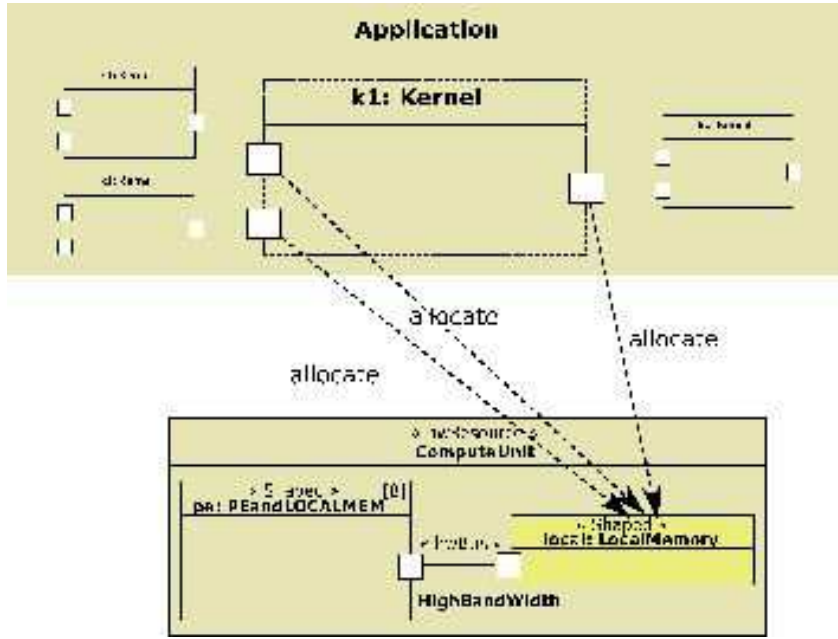}
\caption{Flowport Assotiation}
\label{assoc}
\end{figure}

By using UML links we can associate \textit{flowPorts} to memories in architecture models. Each port has attributes and associations that permit us to define size and data type for example. Thus, developers can define their application models where the data will be stored and how much size the data will take. In the fig. \ref{assoc} we can see a simple example of allocation. The ports of the \textit{k} instance (from Kernel application component) are allocated in \textit{local} instance (from LocalMemory architecture component). Then, using the \textit{memory mapping} transformation (that is part of the chain of transformations applied to the input model), we create a structured element tree having all information about size, type and, mainly, where data will be placed in a spatial or temporal way.

\section{Environment and Case Study}
The architecture metamodel for GPUs is used within the transformation chain in order to create an hybrid model which is closer to the target code. EMF is the Eclipse Modeling Framework used by IBM's open source Eclipse project \cite{eclipse10}. We used Eclipse platform to specify the memory mapping metamodel and models for GPU architecture used in this work. The DaRT team has made a new MARTE metamodel based on MARTE profile for UML. This EMF metamodel provides classes and other elements that allow us to create models which describe the applications and hardware characteristics, and the relation between them. The main goal is to define the maximum number of information about the application (and its parallelism) and hardware, thus we can create transformation rules for generating models and source code. This environment provides some tools like QVTO and Acceleo. The first one is the language from the Meta Object Facility Query/View/Transformation (MOF QVT) \cite{qvt07} specification. QVT is the solution for model transformations in the OMG modeling framework. This solution is a defacto standard for model transformations. QVT Operational (QVTO) is a completely imperative language, which only supports model transformation scenarios in an unidirectional M-to-N fashion. The second one (Acceleo)\cite{acceleo10} is a non-standard solution for model-to-text(or code) transformation. Both are used in the transformation chain implemented on \textit{Gaspard2} environment, whose the main objective is code generation for embedded systems.\\

\begin{figure}[ht]
\begin{center}
\includegraphics[width=.7\textwidth]{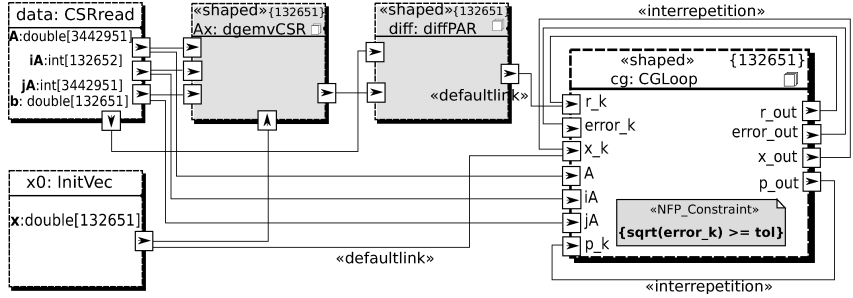}
\end{center}
\caption{ UML/MARTE Model for Setup and CG Overview}
\label{fig-cg}
\end{figure}

\begin{figure}[ht]
\begin{center}
\includegraphics[width=.9\textwidth]{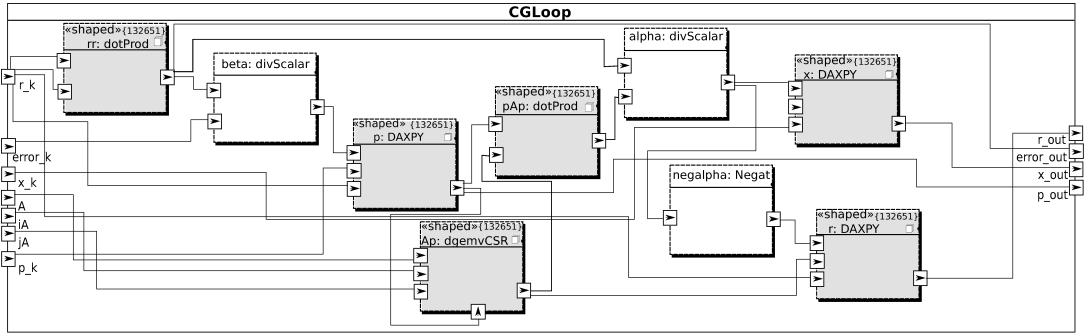}
\end{center}
\caption{Conjugate Gradient UML/MARTE Model}
\label{fig-cgloop}
\end{figure}

The approach presented in this paper was inserted in the Marte to OpenCL transformation chain \cite{rodrigues11}. This chain allows us to create an application and test a generated code. The application in the test environment is the conjugate gradient algorithm implementation. The conjugate gradient (CG) method\cite{golub-vanloan:1996} is often used in modeling and simulation of electrical systems. It should only be applied to systems that are symmetric or Hermitian positive definite. Input data are resulting from a FEM model of an electrical machine. The matrix is stored in \textit{Compressed Sparse Row (CSR)} format having \textit{N}=132651 and \textit{NNZ}=3442951. The CG algorithm is modeled in MARTE as presented in the figure \ref{fig-cg}, where data reading and initial configurations are defined by stereotyped blocks. Highlighted gray blocks represent tasks, which are mapped onto as many devices as we want to distribute the task job. Tasks, such as DGEMV(sparse), are repetitive and, thus, potentially parallel. The CGLoop is a 132651 loop which some of its input data are recovered between continuous iterations. A \textit{continue-condition} is specified by a constraint attached to the CG block, so the loop can stop before running all iterations. The figure \ref{fig-cgloop} is an internal view of the CGLoop modeled in the figure \ref{fig-cg}. Here scalar operations run on CPU processor, and repetitive operations run on GPU processors. Details about deployment of elementary tasks (operations), data and task allocation to architecture, scheduling, grid definition, and so on, can be found in \cite{gaspard2} and they are not discussed in this paper due to scope and space limitation.

We used four double-precision implementation versions of CG. The first one (the reference result) is sequential and uses the Matlab's \textit{pcg} function. The other ones are automatically generated OpenCL implementations whose kernels are launched onto 1, 2 and 4 devices, respectively. The number of used devices depends of the task allocation process. The hardware used is composed by a \textit{2.26GHz Intel Core 2 Duo} processor and S1070 unit (4 Tesla T10 Nvidia GPU). Usually, manually written codes have better performance than automatic ones. However, these automatically generated CG implementations have an expressive performance(table \ref{resultst}) compared to sequential code (time results include just computing and data transfer times in CG loop). The multi-GPU aspect is verified in the two latest versions. The code generation compiler decides equally the task partitioning to the multiple devices. The gain is not linear(though significant) due to extra data transfers among cpu and devices. A detailed analysis about solvers and Multi-GPU can be found in \cite{springerlink:10.1007/s00450-010-0112-6}.

\begin{table}[ht]
	\centering
	\begin{tabular}{|r|llll|}
                \hline
		conjugate gradient & \#iter & time(s) & speed-up & gflops \\
                \hline
		Matlab PCG & 117 & 3.17 & 1 & .303 \\
		OpenCL (1 GPU) & 116 & 0.659 & 4.81 & 1.45 \\
		OpenCL (2 GPU) & 116 & 0.461 & 6.87 & 2.07 \\
		OpenCL (4 GPU) & 116 & 0.380 & 8.34 & 2.50 \\
                \hline
	\end{tabular}
	\caption{Performance Results; N=132651, NNZ=3442951, tol=1e-10}
	\label{resultst}
\end{table}

\section{Conclusions}
The result model based on MARTE profile and metamodel extensions allowed us to describe specifical details of GPUs. Our approach presents a solution to specify a GPU architecture in terms of processor and memory hierarchy validated by a case study. As applications for embedded systems are evolving fast, MDE approaches are more suitable to these systems. Advances in GPU architecture modeling allow developing model-to-model and model-to-code transformations. These transformations contribute to generate executable code for architectures based on GPU.

An important contribution of our approach is in the memory modeling aspect. GPUs have inherent characteristics that require a optimal data allocation from developer. These requirements imply better data alignment and bandwidth use for obtaining application performance. Future works will explore these metamodel and model to generate code more efficient.

\bibliography{sympa2011}

\begin{thebibliography}{10}

\bibitem{marte10}
{Modeling and Analysis of Real-time and Embedded systems (MARTE)}.
\newblock \url{http://www.omgmarte.org/}.

\bibitem{cuda}
{NVIDIA CUDA Compute Unified Device Architecture}.
\newblock \url{http://www.nvidia.com/cuda/}.

\bibitem{MOF}
{OMG's MetaObject Facility}.
\newblock \url{http://www.omg.org/mof/}.

\bibitem{opencl10}
{OpenCL - The open standard for parallel programming of heterogeneous systems}.
\newblock \url{http://www.khronos.org/opencl/}.

\bibitem{Bezivin}
Jean B\'{e}zivin and Olivier Gerb\'{e}.
\newblock Towards a precise definition of the omg/mda framework.
\newblock In {\em Proceedings of the 16th IEEE international conference on
  Automated software engineering}, ASE '01, pages 273--, Washington, DC, USA,
  2001. IEEE Computer Society.

\bibitem{springerlink:10.1007/s00450-010-0112-6}
Ali Cevahir, Akira Nukada, and Satoshi Matsuoka.
\newblock High performance conjugate gradient solver on multi-gpu clusters
  using hypergraph partitioning.
\newblock {\em Computer Science - Research and Development}, 25:83--91, 2010.

\bibitem{eclipse10}
{Eclipse}.
\newblock {\em {Eclipse Modeling Framework}}, 2011.
\newblock \url{http://www.eclipse.org/emf/}.

\bibitem{gaspard2}
A.~Gamati\'{e}, S.~Le Beux, E.~Piel, R.~Ben Atitallah, A.~Etien, P.~Marquet,
  and J-L. Dekeyser.
\newblock A model driven design framework for massively parallel embedded
  systems.
\newblock {\em ACM Transactions on Embedded Computing Systems (TECS). (to
  appear)}, 2011.

\bibitem{golub-vanloan:1996}
Gene~H. Golub and Charles~F. Van~Loan.
\newblock {\em {Matrix Computations}}.
\newblock The Johns Hopkins University Press, 3rd edition, October 1996.

\bibitem{ES}
Holger Lange, Florian Stock, Andreas Koch, and Dietmar Hildenbrand.
\newblock Acceleration and energy efficiency of a geometric algebra computation
  using reconfigurable computers and gpus.
\newblock {\em Field-Programmable Custom Computing Machines, Annual IEEE
  Symposium on}, 0:255--258, 2009.

\bibitem{rodrigues11}
A.~{W}endell O.~{R}odrigues, {F}r{\'e}d{\'e}ric {G}uyomarc'{H}, and
  {J}ean-{L}uc {D}ekeyser.
\newblock {A}n {MDE} {A}pproach for {A}utomatic {C}ode {G}eneration from
  {MARTE} to {O}pen{CL}.
\newblock Technical report, INRIA Lille - RR-7525.
\newblock http://hal.inria.fr/inria-00563411/PDF/RR-7525.pdf/.

\bibitem{acceleo10}
{Obeo}.
\newblock {\em {Acceleo - Model to Text transformation}}, 2011.
\newblock \url{http://www.acceleo.org/}.

\bibitem{qvt07}
{OMG}.
\newblock {\em {M2M/Operational QVT Language}}, 2011.
\newblock \url{http://wiki.eclipse.org/M2M/QVTO/}.

\end{thebibliography}

\end{document}